# A-FC: An Activity-Based Delay Tolerant Routing Protocol for Improving Future School Campus Emergency Communications

Chengjun Jiang
School of Computer Science
University of Nottingham
psxcj11@nottingham.ac.uk

Milena Radenkovic
School of Computer Science
University of Nottingham
milena.radenkovic@nottingham.ac.uk

## Abstract

**School Campus emergency communication systems are vital for safeguarding student safety during sudden disasters such as typhoons, which frequently cause widespread paralysis of communication infrastructure. Traditional Delay-Tolerant Network (DTN) protocols, such as Direct Delivery and First Contact, struggle to maintain reliable connections in such scenarios due to high latency and low delivery rates. This paper proposes the Activity-based First Contact (A-FC) protocol, an innovative routing scheme that leverages real-world social roles to overcome network partitioning by mandatorily uploading messages to highly active "staff nodes". We constructed a real-world evaluation scenario based on the topology of Fuzhou No. 1 Middle School. Simulation results demonstrate that the A-FC protocol significantly outperforms baseline protocols, achieving approximately 68% message delivery probability and reducing average delay to 4311 seconds. With an average hop count of merely 1.68, this protocol establishes a low-cost, highly reliable backup communication model for school campus disaster response.**

## 1. Introduction

### 1.1 Background

In densely populated environments such as school campuses, communication infrastructure constitutes the lifeline for emergency response. Should catastrophic events like typhoon occur, traditional communication methods—such as mobile cellular networks or Wi-Fi—may become paralysed, with grave potential consequences including delayed rescue efforts and compromised search-and-rescue operations. In the coastal city of Fuzhou, China, the threat of powerful typhoons is a frequent occurrence. Following a typhoon's landfall, power and internet outages are commonplace, with many areas becoming inundated. This results in significant economic losses and poses a threat to the lives and safety of the populace.

[10] describe Rescue personnel are undertaking tasks such as draining floodwater and evacuating residents. For instance, as illustrated in Figure 1, Typhoon Doksuri inflicted severe damage upon Fuzhou in 2023.

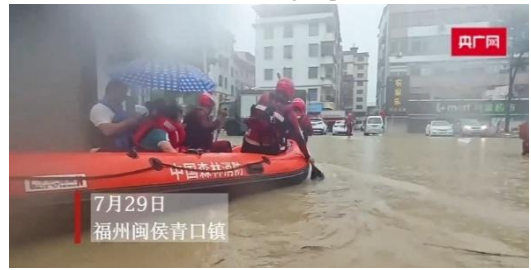

Figure 1 News regarding Typhoon Doksuri[10]

Figure 2 depicts Fuzhou's Olympic Sports Centre, which suffered extensive damage.

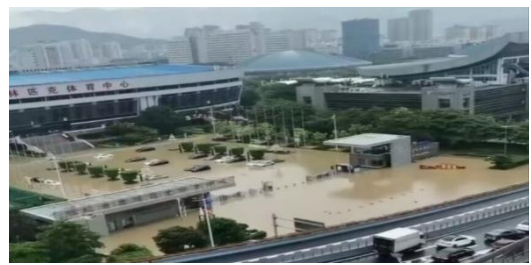

Figure 2 Fuzhou Olympic Sports Centre has been affected by the disaster.[9]

Figure 3 shows the No.1 Middle School of Luoyuan County [14], where torrential rain triggered by the typhoons caused a landslide.

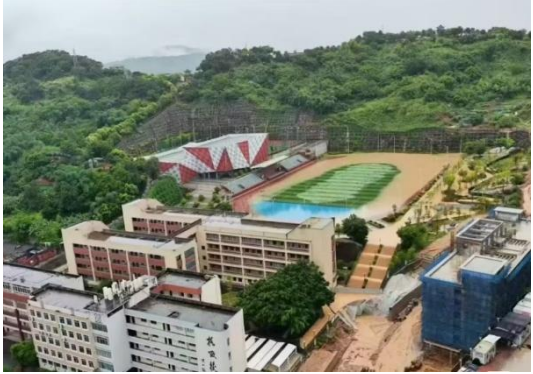

Figure 3 The No.1 Middle School of Luoyuan County [14] has been affected by the disaster.

Figure 4 illustrates the devastation suffered by Fuzhou's historic site "Three Lanes and Seven Alleys" [15] following the passage of Typhoon Doksuri.

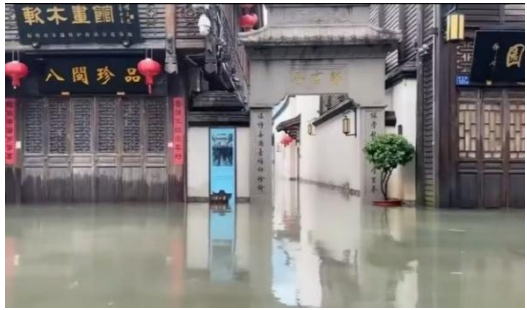

Figure 4 Fuzhou's renowned historic street, "Three Lanes and Seven Alleys" [15] has been affected by the disaster.

Therefore, it is of practical significance to investigate the operation of DTN [1] network protocols in disaster-stricken environments and to develop new protocols. [1] argues that Delay-tolerant network (DTN) [1] architecture is designed for such "challenging networks". It does not rely on end-to-end real-time connections but instead utilises the mobility of nodes (such as students carrying smart communication devices) to transmit information through a "store-carry-forward" approach. [1]

This paper aims to evaluate the fundamental performance of existing protocols and understand their limitations in disaster scenarios.

The research will test the performance of standard DTN protocols [1] during natural disasters such as earthquakes or typhoons, comparing them with the Activity-Based First Contact (A-FC) protocol.

Although benchmark protocols such as Direct Transmission and First Contact [3] are well-established, their efficacy during communication disruptions caused by disasters within school campus environments remains to be validated. This paper seeks to determine the most reliable and suitable school campus emergency communication strategy through comparative analysis, thereby fulfilling the requirement for establishing a highly dependable emergency backup communication system.

## 2. DTN Protocol Descriptions and Novel Activity-Based First Contact Protocol

This section outlines the three protocols employed in this paper, comprising two benchmark protocols and the proposed Activity-Based First Contact (A-FC) protocol.

Direct delivery [3] constitutes the most fundamental transmission method within delay-tolerant networks (DTNs) [1]. Its mechanism is straightforward: the source node stores the message until it enters the direct communication range of the destination node, at which point delivery occurs without utilising any intermediate relay nodes [3]. As noted by Spyropoulos et al., this protocol serves as the baseline for transmission performance [3]. In school campus emergency scenarios with fixed source nodes (e.g., emergency command centres), this protocol demonstrates the necessity of multi-hop routing.

First Contact [3] is a standard single-copy routing protocol that leverages node mobility while maintaining minimal network overhead. "Upon encountering another node not holding the message, the current carrier forwards the message to this new node and subsequently deletes its own copy to ensure only a single copy exists in the network" [3]. Although First Contact [3] efficiently manages communication resources, its forwarding strategy is inherently "blind" [3]. As the protocol's mechanism cannot perceive the receiving node, it often delivers messages to nodes distant from the destination, resulting in high latency.

### 2.1 Activity-Based First Contact (A-FC)

To address the blind spots of the first-contact mechanism, this paper proposes the Activity-Based First Contact (A-FC) mechanism. This design draws inspiration from social-based routing protocols (such as PROPHET [4] and BUBBLE Rap [5]), which utilise node metrics (e.g., forwarding reliability or centrality) for forwarding decisions. Conceptually, A-FC retains the single-copy mechanism of initial contact to ensure zero overhead while introducing an activity-based "activation" forwarding strategy. Here, nodes are categorised as "low activity" (students) or "high activity" (staff), with message forwarding activated only when the receiving node's activity exceeds that of the current carrier.

## 3. Construction of the Hypothetical Topology Fuzhou No.1 Middle School

This hypothetical scenario uses the actual geographical layout of Fuzhou No.1 Middle School, which has been reconstructed to enhance the realism of the experiment.

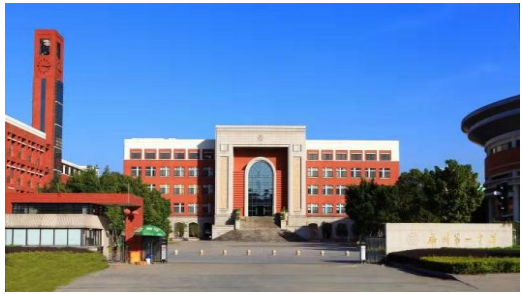

Figure 5 Fuzhou No.1 Middle School Campus Scenery [12]

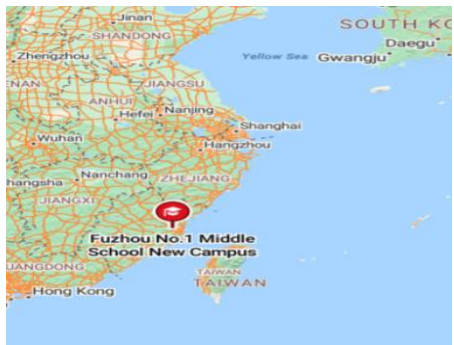

Figure 6 Location of Fuzhou No.1 Middle School [13]

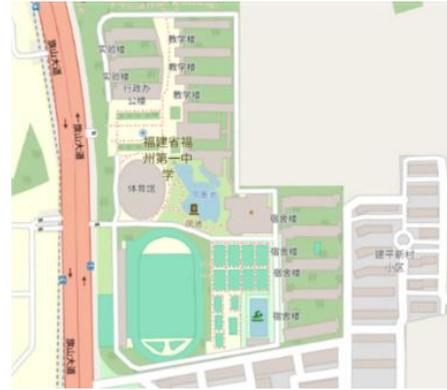

Figure 7 Map of Fuzhou No.1 Middle School [11]

Figure 8 is the map on One Simulator[2]. The simulation area is set at 4500m × 4500m, representing the entire school campus.

The map defines the school campus road network. To ensure the map-based mobility model functions correctly, all path nodes have been processed to guarantee full connectivity. This configuration ensures nodes move along actual school campus roads rather than traversing buildings, thereby simulating realistic evacuation movement.

To evaluate the proposed A-FC protocol in realistic disaster scenarios, we constructed a ONE simulator environment [2] based on the topology of Fuzhou No. 1 Middle School. The simulation involved 120 mobile nodes divided into two independent groups to reflect

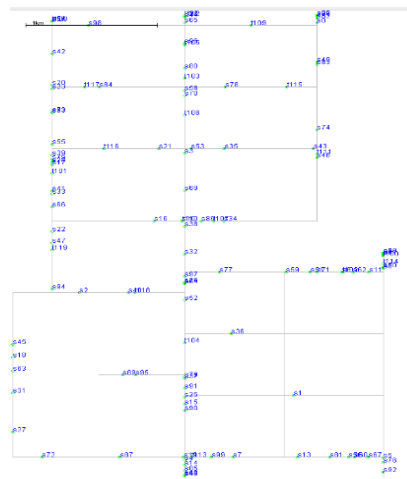

Figure 8 Map of Fuzhou No.1 Middle School in One Simulator

the school campus environment: the student group (s, 100 nodes) and the staff group (t, 20 nodes).

The mobility model employs "ShortestPathMapBasedMovement", ensuring nodes traverse along actual roads rather than through buildings. Staff nodes are configured with higher movement speeds (1.0–2.0 m/s), simulating security personnel or rescue workers acting as "data relays". Student nodes move at slower walking speeds (0.5–1.5 m/s). To simulate a communication black-box scenario with disrupted cellular networks, all nodes communicate via short-range "SimpleBroadcastInterface" (emulating Bluetooth) with a range of 10 metres and a transmission rate of 2 Mbps.

Traffic generation followed a high-load emergency pattern: messages ranging from 500KB to 1MB in size (simulating evacuation maps or voice instructions) were generated every 25 to 35 seconds, with a 300-minute time-to-live (TTL) to ensure message validity throughout the critical rescue window. Detailed simulation parameters are presented in Table 1.

| Parameter | Value |
| --- | --- |
| Simulator | The ONE Simulator |
| Simulation Area | 4500m x 4500m |
| Total Nodes | 120 |
| Node Groups | Students (100), Staff (20) |
| Mobility Model | ShortestPathMapBasedMovement |
| Velocity | Students: 0.5–1.5m/s  Staff: 1.0–2.0 m/s |
| Interface | Bluetooth (10m range, 2 Mbps) |
| Message Size | 500 kB – 1 MB |
| Generation Interval | 25 – 35 s |
| time-to-live (TTL) | 300 minutes |

Table 1 Simulation Parameters

**3.1 Hypothetical Scenario Design and Setup of Fuzhou No.1 Middle School after Typhoon Disaster**

To evaluate the robustness of the A-FC protocol under extreme conditions, this experiment assumes a specific disaster scenario:

Crisis Event: A super typhoon makes landfall in Fuzhou, causing severe damage to the power and network infrastructure within the Fuzhou No. 1 Middle School campus. The entire school campus is plunged into a "communication blackout".

Emergency Task: The temporary rescue command centre (comprising school campus staff and security personnel) located at the south-west corner playground urgently needs to transmit "emergency evacuation routes", and "designated assembly points" to students trapped in the south-east corner dormitory area and the north teaching block.

Communication Challenge: With base stations rendered inoperable by the typhoon, communication relies solely on a temporary ad-hoc network established via Bluetooth (short-range, low-power) on students and staff members mobile phones. The critical challenge lies in swiftly and accurately delivering information to every student across the school campus, despite limited battery life and restricted movement.

Figure 9 shows the initial state of the emergency incident. The two gymnasiums on the left side of the central thoroughfare serve as the rescue command centre due to their dense node concentration and sufficient area, yet numerous students remain trapped in peripheral zones. For instance, student S65 is situated in the upper-right quadrant (teaching area) and has lost contact with teaching staff. Owing to typhoon-induced poor visibility and road destruction, S65 will be entirely isolated.

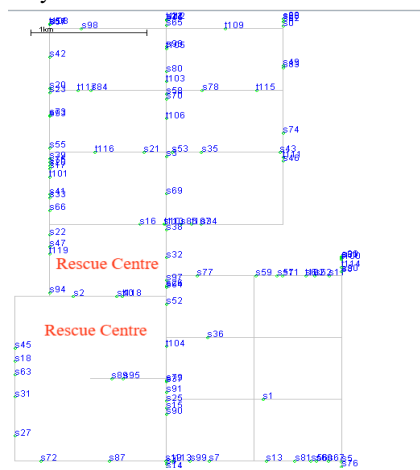

Figure 9 Initial state of the emergency incident.

Figure 10 shows student S65 attempts to move within the teaching zone to seek rescue. However, constrained by the Bluetooth transmission range (10 metres) and the absence of other students nearby to receive messages, S65 remains disconnected.

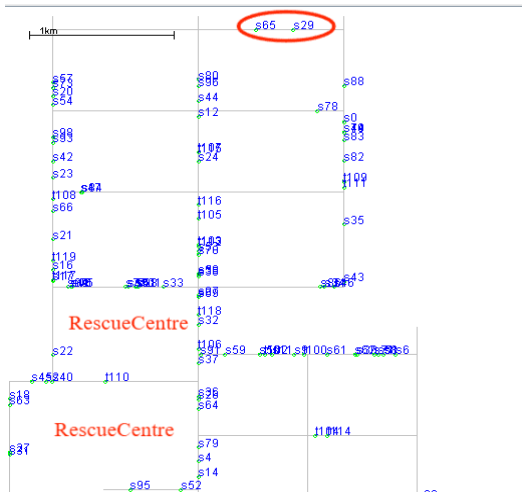

Figure 10 Student S65 seeking assistance

Figure 11 shows successful relay achieved via the A-FC protocol. Student node S65 encounters the highly active teacher node t111. Under the A-FC protocol, the message is successfully uploaded to t111. This teacher then transmits evacuation information and routes to student S65. Should the student require assistance, help is provided; if the student is in good condition, they continue searching and disseminating evacuation information.

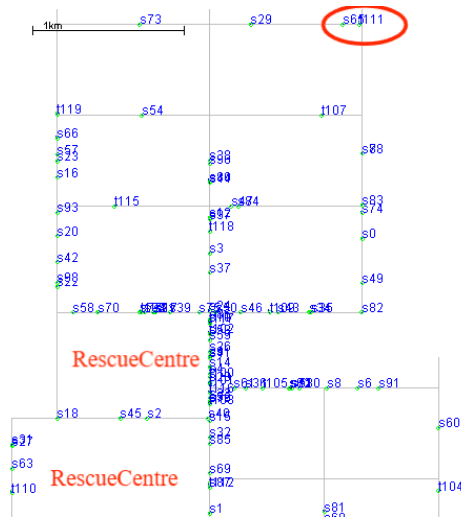

Figure 11 Student S65 receives information from teacher t111 and is assisted.

## 4. Multi Criteria Multi-Dimensional Results Analysis

### 4.1 Performance Metrics and Comparative Data

We compared the performance of the proposed A-FC protocol against two protocols: Direct Delivery [3] and First Contact [3]. The comparative analysis focused on three key metrics: delivery probability, average delay, and hop count. Simulation results are summarised in Table 2.

| Protocol | Delivery Prob. | Avg. Latency | Avg. Hop Count |
|---|---|---|---|
| A-FC | 0.6797 | 4311.7 | 1.6813 |
| Direct Delivery | 0.4743 | 6010.4 | 1.0000 |
| First Contact | 0.2433 | 4641.8 s | 18.8397 |

Table 2: Comparison of Key Performance Metrics for Three Routing Protocols

Figure 12 and Table 2 demonstrate that the A-FC protocol exhibits exceptional reliability, achieving a delivery probability of approximately 68%. This performance advantage stems from the protocol's socially aware design, which utilises highly active employee nodes as mobile relays. These "super carriers" effectively traverse network partitions, transmitting evacuation instructions to isolated student groups. By contrast, the First Contact protocol [3] performed poorly, achieving the lowest delivery rate (approximately 24%). Its blind forwarding strategy caused messages to circulate within densely populated areas (such as dormitory blocks) without reaching the rescue centre. The Direct Delivery protocol [3] demonstrated moderate performance (approximately 47%), but its reliance on direct source-destination encounters prevented coverage of larger areas.

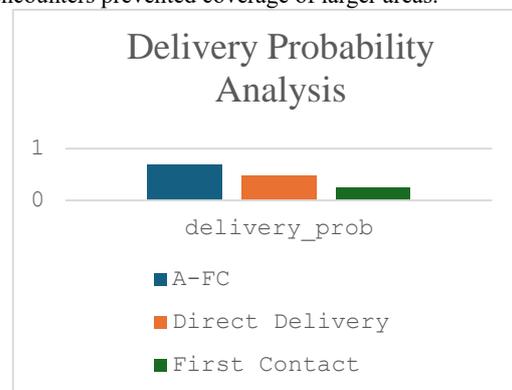

Figure 12 Delivery Probability Analysis

During the "golden hour" of rescue operations, speed is paramount. Figure 13 demonstrates that A-FC exhibits the lowest average delay, at approximately 72 minutes. By uploading messages to faster-moving staff nodes, A-FC significantly reduces the time messages spend in the buffer. In contrast, Direct Delivery [3] suffers from message delays exceeding approximately 100 minutes due to slower-moving student nodes and limited communication range.

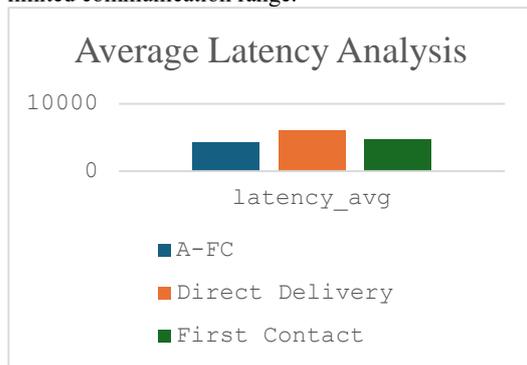

Figure 13 Average Latency

In terms of resource consumption, Figure 14 highlights the efficiency of the A-FC mechanism. With an average hop count of merely 1.68, A-FC confirms that the vast majority of messages follow the optimised "student → faculty → destination" path. This stands in stark contrast to First Contact [3], which exhibits an average hop count of 18.84. Such a substantial number of relay nodes indicates the presence of "hot potato" forwarding behaviour, which not only causes network congestion but also rapidly depletes the battery life of mobile devices – a critical issue in disaster scenarios.

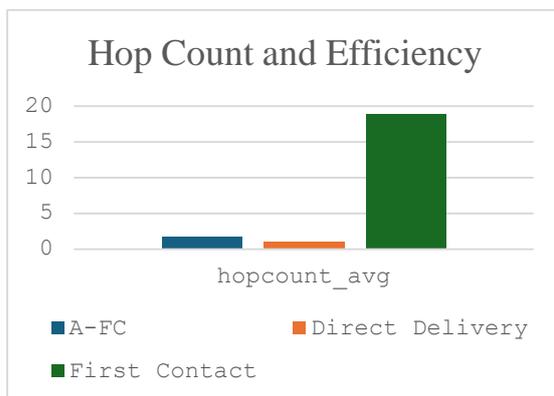

Figure 14 Hop Count and Efficiency

## 5. Conclusion and Future Work

This paper addresses emergency communication requirements within school campus environments during sudden disasters (such as typhoons causing communication infrastructure failure).

The A-FC protocol proposed in this paper introduces real-world social roles, mandating message uploads to highly active "staff nodes". Experimental results demonstrate this strategy significantly boosts message delivery rates to approximately 68% while reducing average latency to 4311 seconds. With an average hop count of merely 1.68, A-FC successfully establishes an efficient emergency information transmission model.

This paper confirms that leveraging inherent social roles within school campuses to optimise routing choices proves exceptionally effective during extreme disasters where communication blackouts occur.

By contrast, reliance on conventional protocols could result in missing the critical rescue window. A-FC offers a highly practical solution for constructing low-cost, high-reliability backup emergency communication systems for school campuses.

Although the A-FC protocol performed well in this experiment (simulation at Fuzhou No. 1 Middle School), there remains scope for optimisation. Future work may further refine this protocol by integrating advanced protocols evaluated in key literature.

Congestion control is of paramount importance. Future integration of the CafRep[6] mechanism is feasible; this utility-driven framework dynamically redirects traffic from congested nodes to less congested areas, thereby enhancing overall network stability and throughput.

Concurrently, research into adaptive caching mechanisms may be pursued. The CafRepCache[7] protocol introduces a collaborative cognitive caching strategy, optimising storage resource utilisation by predicting node and content importance.

Moreover, leveraging CognitiveCache[8], the emergency network could employ Multi-Agent Deep Reinforcement Learning (MADRL) to autonomously learn and adjust forwarding strategies within complex dynamic environments.

Finally, battery capacity constraints will be incorporated into future simulation to evaluate the protocol's endurance performance under realistic energy limitations.

# Reference

[1] K. Fall, "A delay-tolerant network architecture for challenged internets," in *Proc. ACM SIGCOMM '03*, Karlsruhe, Germany, Aug. 2003, pp. 27–34.

[2] A. Keränen, J. Ott, and T. Kärkkäinen, "The ONE simulator for DTN protocol evaluation," in *Proc. 2nd International Conference on Simulation Tools and Techniques (SIMUTools '09)*, Rome, Italy, Mar. 2009, pp. 1–10.

[3] T. Spyropoulos, K. Psounis, and C. S. Raghavendra, "Spray and wait: an efficient routing scheme for intermittently connected mobile networks," in *Proc. ACM SIGCOMM Workshop on Delay-Tolerant Networking (WDTN '05)*, Philadelphia, PA, USA, Aug. 2005, pp. 252–259.

[4] A. Lindgren, A. Doria, and O. Schelén, "Probabilistic routing in intermittently connected networks," *ACM SIGMOBILE Mobile Computing and Communications Review*, vol. 7, no. 3, pp. 45–47, Jul. 2003.

[5] P. Hui, J. Crowcroft, and E. Yoneki, "BUBBLE Rap: Social-based forwarding in delay tolerant networks," in *Proc. 9th ACM International Symposium on Mobile Ad Hoc Networking and Computing (MobiHoc '08)*, Hong Kong, China, May 2008, pp. 241–250.

[6] M. Radenkovic and A. Grundy, "Framework for Utility Driven Congestion Control in Delay Tolerant Opportunistic Networks," in *Proc. 7th International Wireless Communications and Mobile Computing Conference (IWCMC)*, Istanbul, Turkey, Jul. 2011, pp. 448–454.

[7] M. Radenkovic and A. Grundy, "Efficient and adaptive congestion control for heterogeneous delay-tolerant networks," *Ad Hoc Networks*, vol. 10, no. 7, pp. 1322–1345, 2012

[8] M. Radenkovic and V. S. H. Huynh, "Cognitive caching at the edges for mobile social community networks: A multi-agent deep reinforcement learning approach," *IEEE Access*, vol. 8, pp. 179561–179574, 2020.

[9] Google Maps, "Fuzhou Olympic Sports Centre, Fujian Province, China"

[10] W. Gong, "Live coverage: 'Doksuri' causes disasters in many parts of Fuzhou; Forest fire brigades carry out drainage and rescue operations," *CNR (China National Radio)*, Jul. 29, 2023

[11] OpenStreetMap "Fuzhou No.1 Middle School"

[12] Fuzhou No.1 Middle School, Campus Scenery

[13] Bing Map, "Fuzhou No.1 Middle School"

[14] Google Maps, "The No.1 Middle School of Luoyuan County"

[15] Google Maps, "Nanhou Street, Dongjiekou Commercial District, Gulou District, Fuzhou City, Fujian Province, China" , "Three Lanes and Seven Alleys"